\documentclass{PoS}

\usepackage{wrapfig}

\title{Status of the Medium-Sized Telescope for the Cherenkov Telescope Array}

\ShortTitle{Status of the MST for the CTA}

\author{\speaker{Markus Garczarczyk}$^{1}$, Stefan Schlenstedt$^{1}$, Louise Oakes$^{2}$, Ullrich Schwanke$^{2}$ 
and the MST team for the CTA consortium\footnote{Full consortium list at http://cta-observatory.org} \\

E-mail: \email{markus.garczarczyk@desy.de}\\ \\
\llap{$^{1}$} Deutsches Elektronen Synchrotron (DESY), Platanenallee 6, D-15738 Zeuthen, Germany \\
\llap{$^{2}$} Humbold University Berlin, Newtonstr. 15, D-12489 Berlin, Germany\\ }

\abstract{The Cherenkov Telescope Array (CTA), is an international project for the next generation ground-based observatory for gamma-ray astronomy in the energy range from 20~GeV to 300~TeV. The sensitivity in the core energy range will be dominated by up to 40 Medium-Sized Telescopes (MSTs). The MSTs, of Davies-Cotton type with a 12~m diameter reflector are currently in the prototype phase. A full-size mechanical telescope structure has been assembled in Berlin. The telescope is partially equipped with different mirror prototypes, which are currently being tested and evaluated for performances characteristics. A report concentrating on the details of the telescope structure, the drive assemblies and the optics of the MST prototype will be given.}

\FullConference{The 34th International Cosmic Ray Conference,\\
30 July- 6 August, 2015\\
The Hague, The Netherlands}

\begin{document}

\section{Introduction}

\begin{figure}[h] \centering
\includegraphics[trim = 0mm 3mm 0mm 5mm, clip, width=1.0\textwidth]{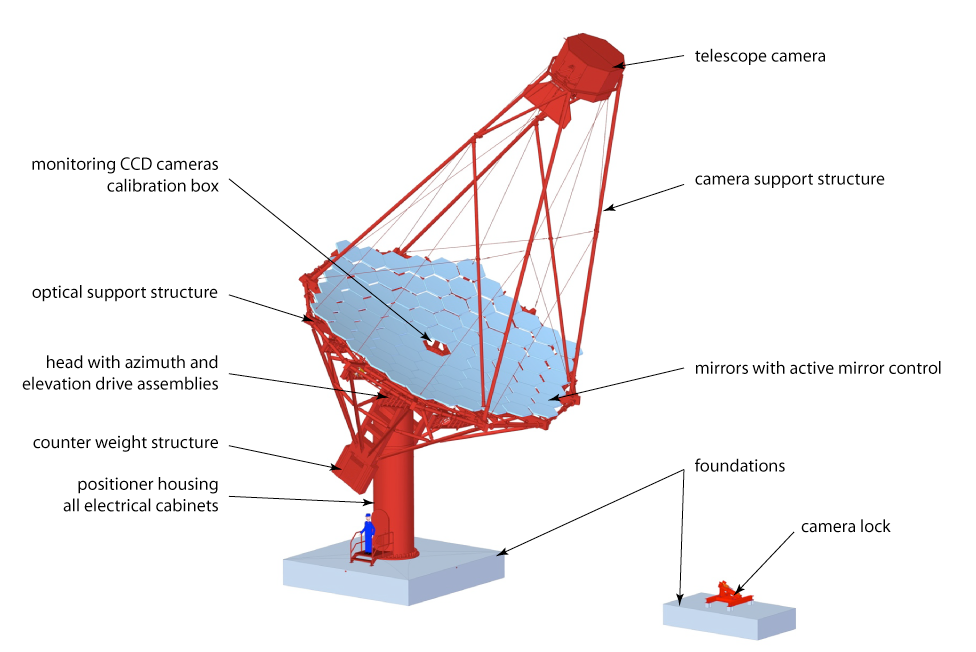}\caption{Illustrative description of the MST design with the main components.}
\label{fig:mst} \end{figure}

The MST is an imaging atmospheric Cherenkov telescope. The telescope optics is based on a modified Davies-Cotton layout with a reflector diameter of 12~m, mirror focal length of 16~m and dish radius of curvature of 19.2~m. Figure~\ref{fig:mst} shows the telescope and indicates its main components.

The telescope mechanical assembly is sub-divided into the following components: Positioner consisting of the tower, head and yokes; optical support structure made out of the dish structure, counter weight structure, mirror access platforms and camera support structure; the azimuth and elevation drive assemblies; camera maintenance structure and foundations.

Two camera concepts are under development for the MST. The design and performance of the camera assemblies FlashCam and NectarCAM are described in~\cite{FlashCam-ICRC, NectarCAM-ICRC} and are not part of these proceedings. The mechanical and electrical interfaces of the cameras to the MST assemblies are identical. The design goal is to allow substitution of one camera type with another camera type on a telescope without major efforts.

\section{Telescope mechanical assembly}

\subsection{Positioner}

The tower of the MST has a cylindrical shape with a diameter of 1.8~m and a wall thickness of 20~mm. The azimuth bearing, installed on top of the tower, acts as the interface to the head and allows rotation along the horizontal axis. The interface of the tower to the foundation is achieved with a flange with pre-stressed screws. The screwed interface allows adjustment of the tower tilt during the assembly process and gives the opportunity to compensate fabrication tolerances of the tower foundation. Access to the tower is possible through a door at the base. All electrical cabinets are stored in the interior of the tower, distributed over three floors. The power and data cables, supplied by the observatory infrastructure, enter the tower through the foundation and are integrated into the main electrical cabinet. Cables connecting the telescope camera and the devices installed on the optical support structure enter the tower through a cable duct on the head. The cables guided through the head are arranged in a way that turning of the telescope azimuth axis by $\pm$270$^{\circ}$ is possible. High ambient temperatures and strong sun radiation during the summer time heats up the interior of the tower. A combination of ventilation and air conditioning system is used to keep the temperature within a safe operation range.

\subsection{Drive assemblies}

The selection of the drive assembly components was optimised to meet the MST requirements to reach any object in the sky above 30$^{\circ}$ in elevation in less than 90~s and to achieve a good running smoothness during typical tracking velocities. The operation range of the azimuth axis was selected to be $\pm$270$^{\circ}$, the elevation axis operation range is 115$^{\circ}$.

\begin{wrapfigure}{r}{0.35\textwidth} \vspace{-4mm} \centering
\includegraphics[width=0.35\textwidth]{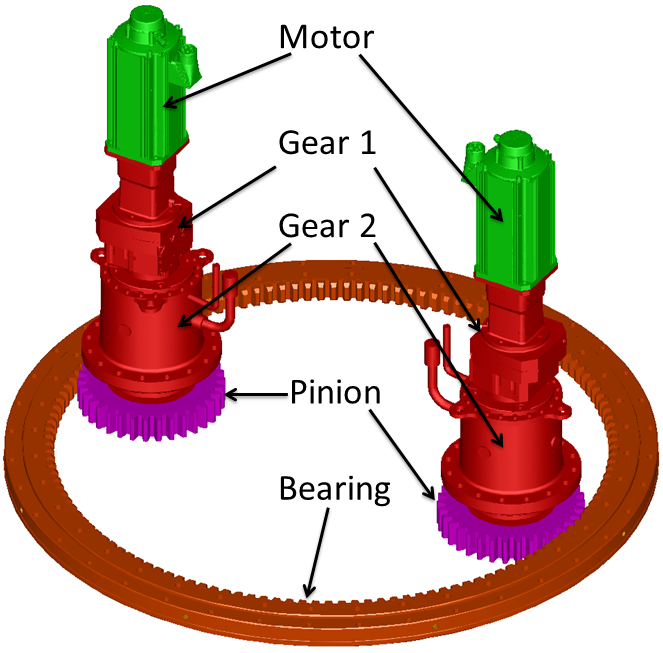}
\caption{Azimuth drive sub-assembly.} \label{fig:mst_azdrive}
\vspace{-30mm} \end{wrapfigure}

The main components of the azimuth drive sub-assembly are shown in Figure~\ref{fig:mst_azdrive} and are as follows:

\begin{itemize}
\item A commercial (Rothe Erde\textregistered) ball bearing with an outer diameter of 1.90~m, 96~mm height and internal geared ring with 136 teeth on a pitch diameter of $\sim$1.6~mm.
\item A gear combination of two standard industry gears (helical and planetary) from SEW.
\item Two synchronous motors from Bosch-Rexroth.
\item Two multi-turn absolute encoders (Heidenhain) resulting in 20" accuracy. 
\end{itemize}

\begin{wrapfigure}{r}{0.35\textwidth} \vspace{0mm} \centering
\includegraphics[width=0.35\textwidth]{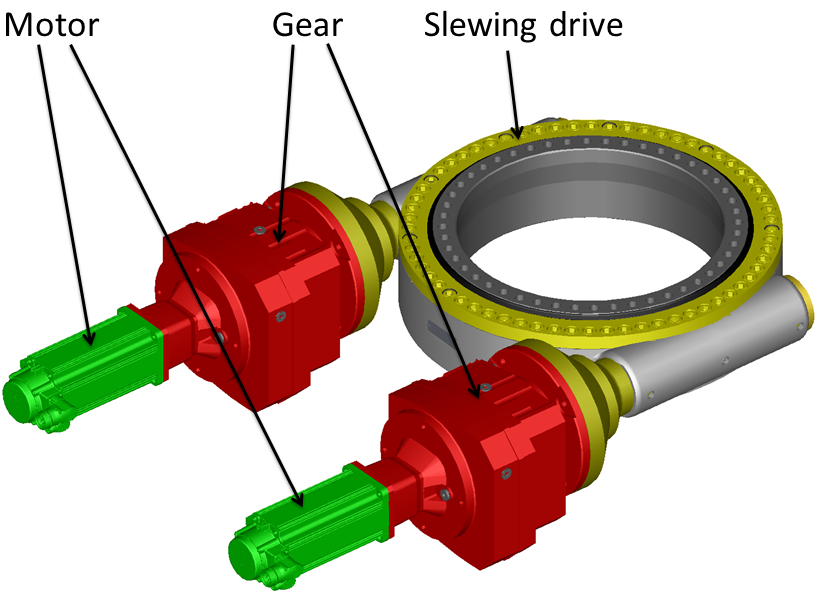}
\caption{One of the two elevation drive sub-assemblies.} \label{fig:mst_eledrivecomp}
\vspace{0mm} \end{wrapfigure}

The elevation drive sub-assembly connects the head with the yokes and provides the rotation of the optical support structure in the vertical direction. The elevation drive is split into two separate sub-assemblies mounted on the left and right side of the head. The main components of the elevation drive sub-assembly are shown in Figure~\ref{fig:mst_eledrivecomp}. Each sub-assembly contains a IMO slewing drive, two gears and two motors. Similar to the azimuth drive sub-assembly, the elevation drive sub-assembly uses two motors per slewing drive resulting in higher safety factor, accuracy and flexibility. Two single-turn absolute encoders with an accuracy of 20" are mounted on each side of the elevation axis.

There are two excitation sources, which can lead to low-frequency oscillations of the telescope structure: External random excitations caused by wind gusts or excitations resulting from the mechanical design, e.g. flexibility of parts with limited stiffness as gears and coupling connections. Both low-frequency oscillations affect the pointing accuracy of the telescope and can be reduced using an active vibration damping control. The oscillations of the camera support structure are measured with an accelerometer sensor, which is connected to the drive control unit. Realtime correction of the motor velocity is fed back into the drive control loop and dampens actively the oscillations. 

To minimise the regular maintenance time a central lubrication assembly is used. Moving components as the bearing, its teeth and slewing drives are greased automatically following the manufacturer recommendations.

\section{Optical assembly}

The MST dish structure is designed to host up to 90 hexagonal mirror segments with a flat-to-flat side length of 1.2~m. In order to fulfil the effective mirror area requirement of $>88$ m$^{2}$ the shadowing of the camera support structure and camera housing was taken into account, this results in 86 mirror segments being required. The individual mirror segments are aligned using an active mirror control to form a uniform reflector.

\subsection{Mirrors}

Three different mirror types are under investigation for the MST. The different technologies provided by the MST stakeholder institutes INAF, CEA and IFJ-PAN are summarised in Table~\ref{tab:mirrors}. 

\begin{table}[h] \centering
\includegraphics[trim = 19mm 222mm 20mm 25mm, clip, width=0.94\textwidth]{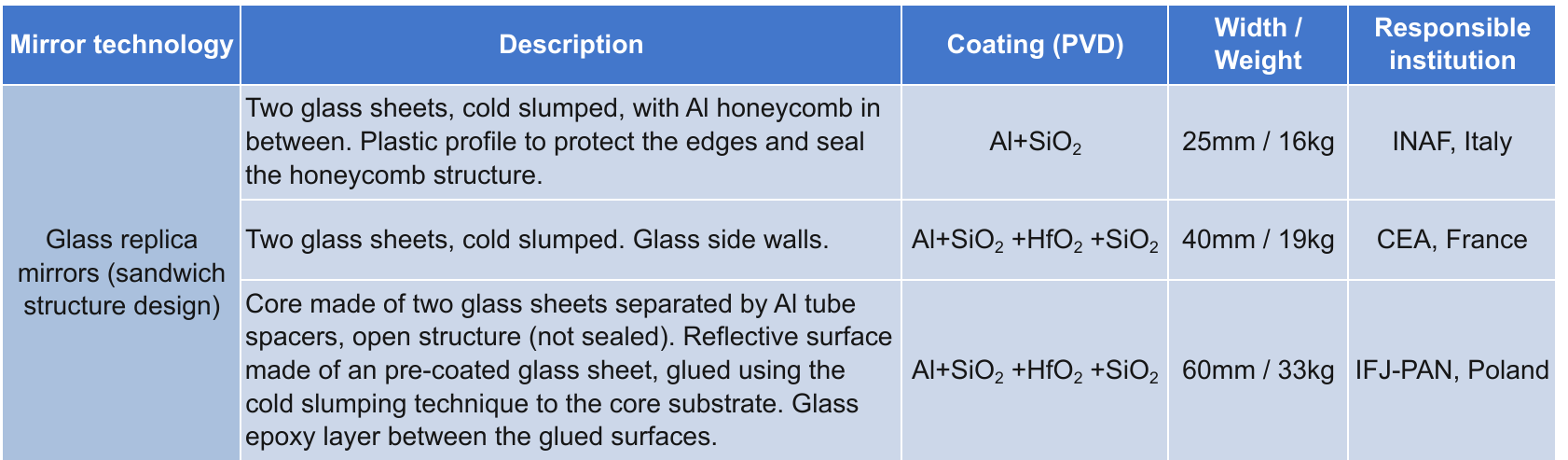}\caption{Summary of MST mirror technologies.}
\label{tab:mirrors}
\end{table}

The mirrors are produced using the cold slumping technique. The replication technique allows high production rate and reproducible optical characteristics. The mirror shape is produced by bending of a thin glass sheet on a precisely formed mould. The smoothness of the surface and therefore the quality of focused spot is attributed by the thickness and quality of the glass sheet as well as precision of the mould's shape. The reflective layer is made of aluminium coating, protected with a SiO$_{2}$ quartz layer / multilayer. Once the design and production line is validated, it is relatively easy to start mass production and produce large quantity of mirrors with exactly the same characteristics. The production rate is determined by the curing time of the glue, e.g. the time the substrate has to be attached to the mould. The production rate of one mirror / day can be scaled with a relatively small investment in a larger number of moulds.

The INAF~\cite{INAF} and CEA~\cite{CEA} mirrors are made with a closed mirror core, the IFJ-PAN mirrors have an open structure core~\cite{IFJ}. The open structure design eliminates the possibility of water accumulation inside the mirror core and consequently its damage at temperatures below the freezing point.

\subsection{Active mirror control}

\begin{wrapfigure}{r}{0.35\textwidth} \vspace{-5mm} \centering
\includegraphics[trim = 0mm 15mm 0mm 0mm, clip, width=0.35\textwidth]{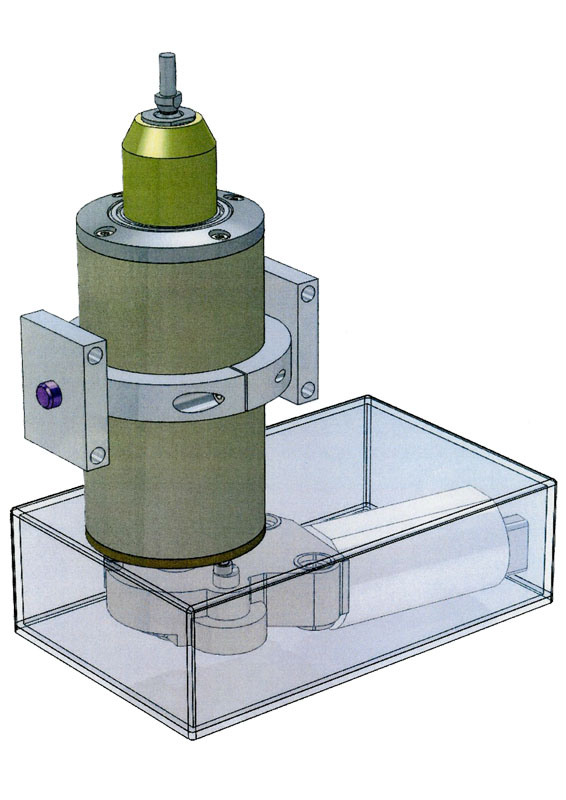}
\caption{T\"ubingen actuator design.} \label{fig:amc} \end{wrapfigure}

The optical support structure of the MST is relatively stiff. Its deformation under gravitational loads and wind impacts at varying elevation angles degrades the overall reflector PSF only marginally. In general, there is no need to readjust the individual mirrors for different elevation angles during normal operation. However, for the initial mirror alignment as well as during any future mirror replacements, e.g. re-coating, it is of advantage, especially from a human safety point of view, if the mirror orientation can be controlled remotely. The AMC can also be used to adjust the focal distance of the telescope reflector, e.g. adapt it to the elevation dependent distance of the shower core maximum. Finally, in an emergency case, e.g. mechanical failure of the drive assembly during the night, the AMC offers a possibility to defocus the reflector in a controlled manner during daytime and reduce the danger caused by reflections of the Sun on the telescope structure and surrounding area. For the above mentioned reasons it is foreseen to equip the MST with AMC.

The T\"ubingen actuator design was selected for the MST~\cite{amc}. The design was initially developed by the Max-Planck-Institut f\"ur Kernphysik for the 28 m diameter H.E.S.S.-II telescope. The mechanical parts were revised by an engineering bureau aiming at improving the design for manufacturability and coping with the MST requirements. The mechanical design of the actuator is shown in Figure~\ref{fig:amc} and is based on:

\begin{itemize}
\item A commercial electric servo motor housed in an aluminium watertight housing. The box is equipped with a venting sintered bronze valve allowing the equalisation of pressure inside the housing with the environment and letting out condensed water vapour (if present).
\item A steel spindle anchored to the motor through a gear and housed in a stainless steel hollow cylinder, to which it is connected by a bronze thread.
\item The stainless steel hollow cylinder housed in another aluminium cylinder, screwed to the actuator box. The two cylinders can freely shift longitudinally with respect to each other, rotation is prevented by two small steel guide poles fixed to the inner cylinder and entering the external one.
\item A ball head connecting the top of the stainless steel cylinder to the interface plate used to connect the actuator with the mirror.
\item As the spindle rotates, the inner cylinder moves up/downwards with respect to the external one, making the actuator head move.
\item The actuators are self-locking when unpowered.
\end{itemize}

The total longitudinal movement range of the actuator is 45~mm. This allows a displacement of the reflected spot at the focal plane by $\sim$8$^{\circ}$ and the possibility, in emergency case, to diffuse reflections of the Sun during daytime from the camera housing. The position of the piston is determined by counting Hall-sensor signals from the motor with respect to an arbitrary reference point. The step length corresponding to a single Hall-count is given by the ratio of the spindle thread spacing (2~mm) and the number of motor Hall-counts for one spindle turn (420) and is in the present version equal to 4.8$~\mu m$. The actuator is controlled by an Mirror Control Unit (MCU) based on an Atmel AT90CANxx\textregistered~CAN micro-controller, which is located inside the actuator box. The electronic board controls the motor movement and checks parameters like motor current consumption, voltage and temperature in order to monitor the health status of the actuator. All MCUs are serially connected to an embedded computer.


\section{Auxiliary assemblies}

\subsection{Telescope pointing calibration}

Accurate calibration of telescope pointing requires precise knowledge of the true pointing direction of the telescope as well as the input (Alt, Az) drive assembly coordinates and the target pointing direction. Corrections to the intended pointing direction due to local and instrumental conditions are calculated offline and stored in a database or in a look-up table. This generally relies on analysing optical sky images taken with one or more CCD cameras mounted on the telescope during dedicated pointing runs.

\begin{wrapfigure}{r}{0.40\textwidth} \vspace{-5mm} \centering
\includegraphics[trim = 25mm 15mm 0mm 0mm, clip, width=0.40\textwidth]{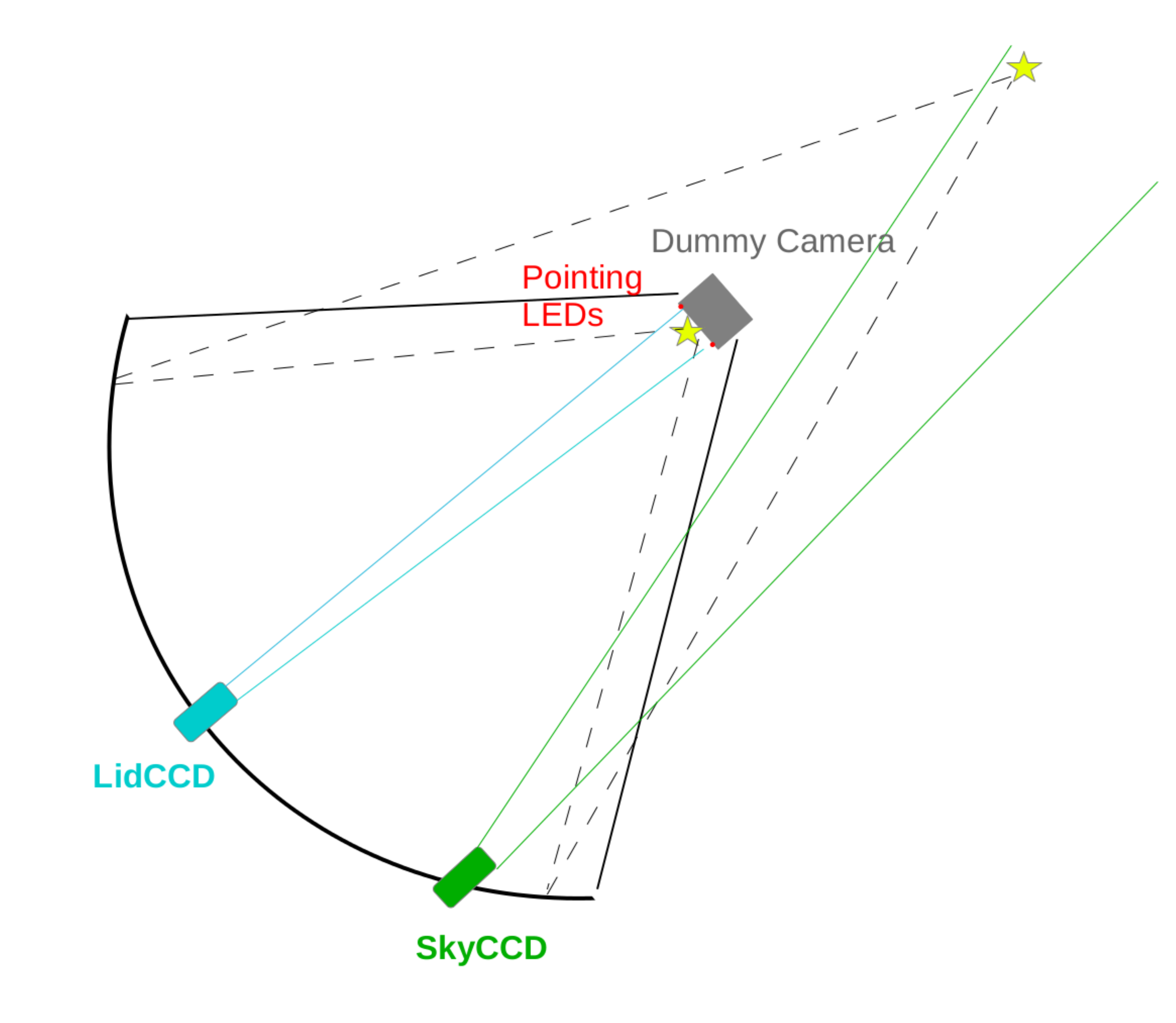}
\caption{Sketch of the 2 CCD camera pointing setup.} \label{fig:mst_cameraprinciple}
\vspace{-2mm} \end{wrapfigure}

Three possible CCD camera configurations are under investigation for the MST pointing calibration. The baseline method is the two CCD camera pointing technique, which has been shown to work successfully on the H.E.S.S. project and is currently being tested on the MST prototype. The method consists of a LidCCD camera installed in the centre of the dish, focused on the telescope camera lid, and a second SkyCCD camera , offset from the optical axis of the OSS with a clear view of the sky. The LidCCD is aligned to the telescope optical axis using the position of dedicated LEDs installed on the telescope camera housing. Other methods, with possible significant gains in precision, use either a single CCD camera with a large enough FoV, such that parts of the telescope camera and the sky are seen simultaneously (reducing the need and possible errors resulting from the mount and alignment of the two CCD cameras), or two CCD cameras installed on one common mount close to each other and placed in the centre of the dish (reducing the effect of camera misalignment due to structure distortions).

\subsection{Condition monitoring}

The MST is equipped with condition monitoring of the telescope structure and drive assembly in order to allow advance detection of possible failures by changes of their performance during its lifetime.

The telescope structure condition monitoring is measuring changes of eigenfrequencies and their amplitudes over time. Long term evolution of these parameters indirect detection of settlements, bending, loosening of bolted/screwed joints or material fatigues. The selected dual-axis sensors (KAS901-54) measure the acceleration and inclination with a sensitivity of 1.2~V/g. The positions of the sensors on the telescope structure were derived from FEM simulations. In total up to seven sensors will be installed on the head, yoke, main dish beams and the camera frame. The interpretation of the data is based on the correlation of each individual sensor with external parameters like wind speed, wind impact direction, telescope acceleration, movement speeds and the telescope position. Response changes of the mechanical structure elements are recognised by shifts of the eigenfrequencies and their amplitudes over time. Initial measurements during the commissioning phase will be used as reference for the time analysis.

The most common of drive assembly components are failures due to wear and imbalance, misalignment and free play. Since all types of damage in rotating machines result in impact impulses causing vibrations to spread out on their housings, accelerometer gauges can be used to detect these vibrations. The number of sensors is a trade-off between cost and sensitivity needed to localise the cause of failure. Based on the validation of the method at the prototype telescope, one sensor for each drive sub-assembly mounted on the gear between the motor and pinion is sufficient. The sensitivity of the transducers should be at least 100~mV/g. \newline
The drive assembly condition monitoring requires stable environment conditions with predefined test movements. Continuous monitoring, as done for the telescope structure condition monitoring is therefore not possible. Test runs include predefined movements of each drive sub-assembly, recording simultaneously the signals of the sensors. The electrical signals are read-out by a vibration monitor (PCH~1420 from company PCH Engineering), which is located in the telescope calibration and monitoring cabinet. The data analysis is based on the comparison of the signals with initial calibration and preset limits. The signals are checked online during the measurement and in the case they exceed the limits a traffic-light system forwards the information to the operator. The data is stored in a database and allows more sophisticated trend analysis to be performed offline. Changes in the long term behaviour of the vibrations indicate subtle damage processes. Recognition of these failures can be done in early stage and allows to plan repairs or replacements of broken components well in advance of a critical failure.


\section{Conclusions}

The planned baseline of CTA includes 40 MSTs. 25 MSTs shall be deployed in the southern hemisphere, and 15 MSTs in the northern hemisphere observatory site. CTA is currently approaching the end of its preparatory phase, where prototypes of telescope assemblies were built and tested. The installation of the telescopes will be, in accordance with the global CTA schedule, divided into a pre-production and a production phase. 

A prototype for the mechanical structure, drive assemblies and auxiliary components of the MST was assembled during 2012 in Berlin. The goals of the prototype were to test and optimise the behaviour of the mechanical structure under loads and environmental influences, the behaviour of the drive and safety concepts and last but not least the functionality of the mirror control and alignment. Three different mirror prototype designs for the MST were tested in the CTA mirror test facilities. The majority of these mirrors were mounted on the prototype dish to complete characterising tests. The mirrors were used to verify the optical support structure deformations and tracking/pointing accuracy of the telescope. The activities carried out during the prototyping and commissioning phase included installation, optimisation, maintenance, commissioning, adjustments, calibrations and surveys.

The MST mechanical structure prototype fulfils the design requirements. Given the early start of the prototyping activities optimisations of several components are currently ongoing. The optimisations emerged from modifications implemented to the SCT~\cite{sct} positioner, as well as experience gained during the production, assembly and commissioning of the prototype telescope. Optimisations focus on cost reduction and improvement of the assembly procedure as well as mass production reproducibility of individual components. The optimisation process will finish in time for the internal MST design review and design freeze for the pre-production series.

\section{Acknowledgment}

We gratefully acknowledge support from the agencies and organisations listed under Funding Agencies at this website: http://www.cta-observatory.org/.


\begin{thebibliography}{99}


\bibitem{FlashCam-ICRC} P\"uhlhofer G. et al., \emph{FlashCam: the fully-digital camera for the medium-sized telescopes of the Cherenkov Telescope Array}, ICRC 2015, these proceedings.


\bibitem{NectarCAM-ICRC} Glicenstein J.F. \emph{NectarCAM: a camera for the MST of the Cherenkov Telescope Array}, ICRC 2015, these proceedings.

\bibitem{INAF} Canestrari R. et al., \emph{Cold-shaping of thin glass foils as a method for mirror processing: from basic concepts to mass production of mirrors}, Optical Engineering, Volume 52, id. 051204 (2013)

\bibitem{CEA} Brun P. et al., \emph{Composite mirror facets for ground based gamma ray astronomy}, NIM A, 714, 58-66 (2013) 

\bibitem{IFJ} Micha\l owski J. et al., \emph{Developments of a new mirror technology for the Cherenkov Telescope Array}, ICRC 2015, these proceedings.

\bibitem{amc} Dick J. et al., \emph{Recent developments for the testing of Cherenkov Telescope Array mirrors and actuators in T\"ubingen}, ICRC 2015, these proceedings.


\bibitem{sct} Rousselle J., \emph{A Medium Sized Schwarzschild-Couder Cherenkov Telescope Design Proposed for the Cherenkov Telescope Array}, ICRC 2015, these proceedings.


\end{thebibliography}
\end{document}